# RISK ASSESSMENT AND MITIGATION OF E-SCOOTER CRASHES WITH NATURALISTIC DRIVING DATA

**Avinash Prabu**
**Zhengming Zhang**
**Renran Tian**
**Stanley Chien**
**Lingxi Li**
**Yaobin Chen**
Indiana University-Purdue University Indianapolis
USA

**Rini Sherony**
Collaborative Safety Research Center (CSRC)
Toyota Motor Engineering & Manufacturing North America
USA

## ABSTRACT

Recently, e-scooter-involved crashes have increased significantly but little information is available about the behaviors of on-road e-scooter riders. Most existing e-scooter crash research was based on retrospectively descriptive media reports, emergency room patient records, and crash reports. This paper presents a naturalistic driving study with a focus on e-scooter and vehicle encounters. The goal is to quantitatively measure the behaviors of e-scooter riders in different encounters to help facilitate crash scenario modeling, baseline behavior modeling, and the potential future development of in-vehicle mitigation algorithms. The data was collected using an instrumented vehicle and an e-scooter rider wearable system, respectively. A three-step data analysis process is developed. First, semi-automatic data labeling extracts e-scooter rider images and non-rider human images in similar environments to train an e-scooter-rider classifier. Then, a multi-step scene reconstruction pipeline generates vehicle and e-scooter trajectories in all encounters. The final step is to model e-scooter rider behaviors and e-scooter-vehicle encounter scenarios. A total of 500 vehicle to e-scooter interactions are analyzed. The variables pertaining to the same are also discussed in this paper.

## INTRODUCTION

**Background**
Electric scooters (e-scooters) have quickly spread in many towns and cities in the U.S. and across many countries in the world as a convenient, cost-effective, and clean mobility option. As early as 2017, 85 cities worldwide adopted e-scooter-sharing systems [1]. Soon after, e-scooter systems were more widely used than bike-sharing systems in the U.S., accounting for 63% of a total of 136 million shared micro-mobility trips in 2019 [2]. As one of the micro-mobility options, the primary use of e-scooters is to address the first and last mile problem, fill the gap between the rider's home or destination and public transport stops, or complete short trips [3], and replace ride-hailing services (39%) (Uber, Lyft, taxi), walking (33%), biking (12), bus (7%), or driving (7%).

In a survey of 7,000 people conducted by Populus in July 2018, 70% of the respondents had a favorable view of e-scooters. This result complies with a recent study published in January 2021 [3], in which researchers found that 73% of survey respondents would or may use e-scooters. Based on existing studies [3]-[9], potential factors affecting the attitudes toward the acceptance of e-scooters include gender, age, race, education level, income level, perceived hedonic quality, prior experiences of ride-hailing services, and familiarity with e-scooters. While economic benefits, locational benefits, and sustainability are driving e-scooter usage [4], infrastructure, weather, and safety might be the top concerns [3].

Rental e-scooters can usually run at a maximum speed of about 10~20 mph. In general, e-scooters are suggested to be ridden in bike lanes or share the road with cars. However, people may also ride e-scooters on sidewalks. Currently, e-scooters are required to follow all traffic rules, but there are no specified regulations established for this new type of transportation tool. This poses various safety concerns for both e-scooter riders and other road users.

For every 100k e-scooter trips, it is estimated that 20-25 injuries result in emergency room (ER) visits [10][11]. Based on this ratio, there might be around 17,000 – 21,000 ER-level injuries in 2019 nationwide. By summarizing news reports, a reporter found that 29 people worldwide have been killed and reported in e-scooter crashes from the summer of 2018 to the end of 2019, mostly in the U.S. Though most e-scooter crash victims are e-scooter riders, there are two pedestrians (including a 90-year-old woman) and one cyclist killed during the crashes too [12]. 169 e-scooter-involved crashes were identified between 2017 to 2019 from U.S. media reports [13]. However, media reports may not cover all cases and can be biased. According to [14], 65 e-scooter-related injuries at Virginia Beach were documented by emergency services during the research period. However, none of these cases were reported in the media. So, the number of severe e-scooter crashes is likely much more than the listed numbers. Based on a study in an emergency department [15], the researchers found that among the randomly sampled crashes causing the E.R. visits, the e-scooter-related crashes jumped from 0 to 25% among all the vehicle-related injuries from 2018 to 2019, when all other sources like cars, bicycles, and motorbikes remained unchanged. In an earlier study [16], the researchers also found a significant increase (500%) in e-scooter-related E.R. visits for two consecutive years after the launch of e-scooter rental services.

**Related Studies**
A naturalistic study was conducted to investigate the interaction of e-scooters and the riding environments, like surrounding objects and potholes, via mobile sensing data [17]. The researchers developed a mobile sensing system, including four low-end sensors: (1) GPS, (2) IMU, (3) 2D LiDAR (12m max distance, 5.5 Hz scan frequency), and (4) portable camera. All the sensing data were facing forward. The research used proximity to objects as the measurement of riding circumstances. Instantaneous proximity maps were aggregated for each trip as the riding complexity measurements. Riding complexity, vibration, and velocity were surrogate riding safety metrics. The main goal was to find the relationships between these safety metrics, riding locations, and surface conditions. Due to the data collection and analysis limitations, this research did not focus on risky interactions or e-scooter riders' behaviors. The study was not able to recognize surrounding objects, their movements, and their detailed interactions. In addition, the study did not record the behavior of the e-scooter riders. All data were aggregated to the whole riding duration for a coarse-level data analysis. The hardware also limited the data quality.

Although there has been very limited naturalistic study on e-scooters, there were some studies that were conducted on e-cyclists, mainly in Europe. Using e-bicycles instrumented with GPS, video cameras, IMUs, pedal sensors, and brake force sensors, the study in [18] collected 88 self-reported safety-critical events from 410 total trips (86 hours of data). Then each event was analyzed using video data about the environmental scenario and threats. The results showed that pedestrians (31%), light vehicles (21%), and bicycles (18%) are the main threats to e-cyclists in critical events. By comparing with the baseline data, the only significant factor affecting the existence of critical events is road location. The intersection is associated with more self-reported safety concerns.

**Research Objectives and methodology**
Unlike vehicle crashes, there is no standard crash database available so far for e-scooter-related crashes [13]. The study collected detailed quantitative behavior data in a naturalistic riding setting. In addition, the effects of individual differences are also modeled and studied.

To our best knowledge, there has not been any systematic research that focuses on the moving behavior and crash scenarios related to e-scooters for public use in an open-road environment.

This paper will address the following important issues:

- Baseline moving patterns of e-scooters in diverse road environments and locations.
- Interaction of e-scooter riders with vehicles and other road users in different scenarios.
- The common scenarios for crashes or near-misses involving e-scooter riders at different risk levels.

# DATA COLLECTION

## Data Acquisition Systems

The primary purpose of the data acquisition system is to record real-time vehicle and e-scooter data. A vehicle-based and an e-scooter-based data collection systems were developed for data collection. Three types of sensors were used in vehicle-based and e-Scooter-based data acquisition systems: Cameras for video data, LiDAR for distance and IMU information, and GPS for latitude, longitude, and altitude data. For portability, the e-scooter system uses smaller-sized (and less quality) cameras and a computing platform compared to the vehicle-based system. The data acquisition system in both e-scooter and vehicle-based systems runs on Ubuntu/ROS platform.

**Vehicle-based Data Collection System:** The vehicle-based data collection system includes six FLIR cameras to cover 360-degree angles, one 64-beam 360-degree Ouster LiDAR, a Reach Emlid GPS, and a desktop computer. The complete system was described in detail a published paper [19]. *Figure* 1 shows the vehicle-based data collection system used in this project.

**E-scooter-based Data Collection System:** There are several constraints while choosing sensors and a data collection device to be mounted on an e-Scooter. Primary considerations are the weight and duration of operation of the system. Since the system is designed to collect data for the surrounding objects, the final system consists of three USB cameras, a 3D LiDAR with integrated IMU, an RTK GPS unit, and an NVIDIA Jetson Tx2 development computer board for data collection. A detailed explanation of the e-scooter-based data collection can be found in [20]. Figure 2 shows the e-scooter-based data collection system in operation.

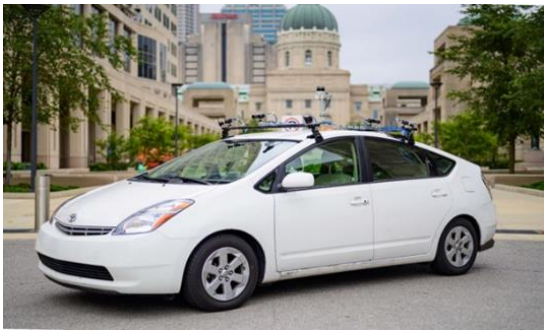
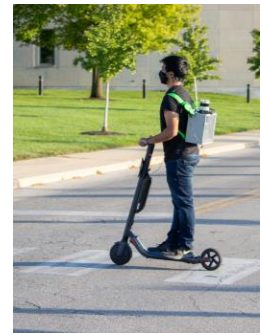

*Figure 1. Vehicle-based Data Collection System.*    *Figure 2. E-scooter-based Data Collection System.*

## Summary of Data Collected – Vehicle-centered Data Collection

68 data collection sessions were completed, with 55 being external subjects. The data collection was mainly performed in Downtown Indianapolis. However, the subjects were given preferred locations to cover during the drive, and the route taken was up to the subject's discretion. Each session was about 20 – 40 minutes in length. The summary of the data collected is as follows:

1. Total time: 35.24 hours –an average of 31 minutes per session.
2. Total data size: ~7.52 TB of raw data.
3. Total number of frames: ~7.6 million (each frame has 6 camera images, LiDAR point cloud, and GPS).
4. Estimated total number of e-scooters: ~400.

## Summary of Data Collected – E-scooter-centered Data Collection

A total of 14 e-scooter-centered data collection sessions were completed. The summary of the data collected is as follows:

1. Total time: 7.5 hours.
2. Total data size: ~1 T.B. of raw data.
3. Total number of frames: ~900,000 million (each frame has 3 camera images, LiDAR point cloud, and GPS).

The estimated nearby cars: >300

# DATA PREPROCESSING

Figure 3 illustrates the data preprocessing pipeline used in this project. A brief explanation of the pipeline is below.

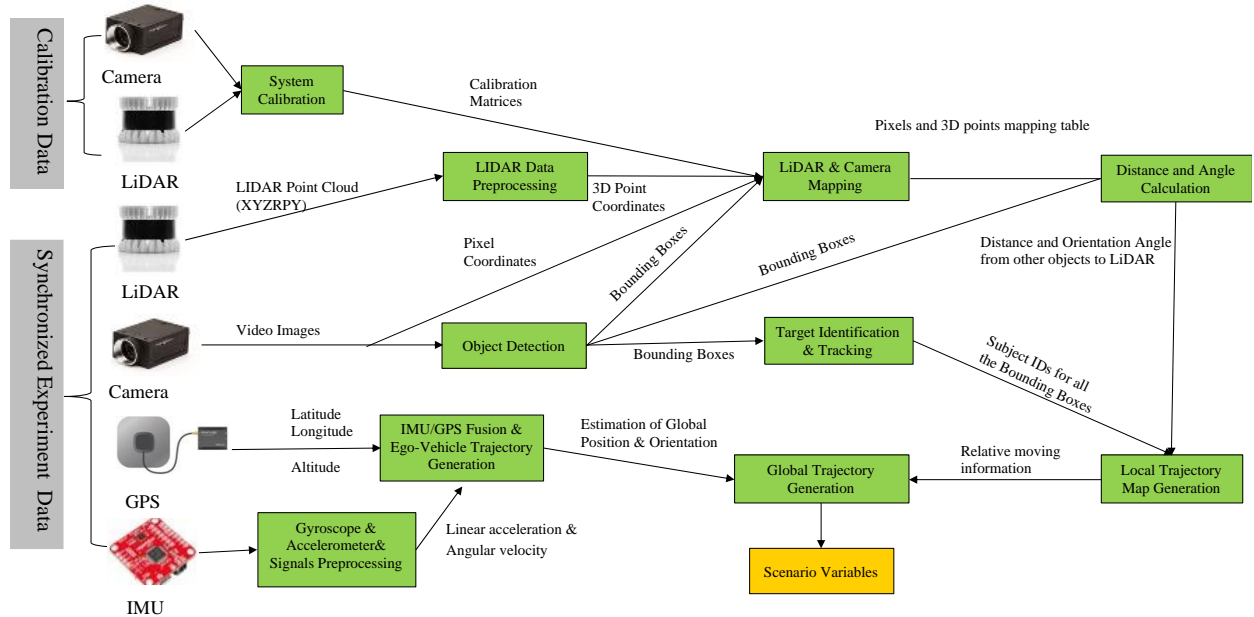

*Figure 3. Data Preprocessing Pipeline.*

1. LiDAR – Camera calibration. The two main components of the calibration are intrinsic and extrinsic matrices. The intrinsic matrix relies purely on the camera and projects 3D points to 2D pixel coordinates [21]. The extrinsic matrix is determined using the Levenberg-Marquardt algorithm and refined for accuracy using the designed user interface [22]. The complete process of system calibration can be found in [20].

2. Automatic object detection and tracking. The fusion output from the system calibration will be used to detect vehicles and humans. From frame to frame, the same vehicle or human will be recognized and connected to track their appearances through video frames. In addition, the 3D coordinates of each detected and tracked object, with respect to the data collection system location, are assigned at each time frame.

    a. A recent algorithm [23] has been successfully implemented to detect and track objects accurately and reliably. This algorithm uses the strategy to detect and track objects simultaneously, which is optimized for processing continued video data. The computational speed of this algorithm is also reasonable for the project at 6 FPS (frame per second).

    b. To detect e-scooter riders automatically, a computer-vision approach is developed in this project. The e-scooter detector structure is shown in Figure 4. This algorithm uses YOLO v3 [24] as the backbone to detect humans. First, the candidate regions on the video images are selected by enlarging the human bounding boxes from YOLO v3, which are then processed with Google MobileNet V2 [25] to obtain visual embeddings. A classifier based on a deep neural network is then trained to classify these humans and surrounding backgrounds into e-scooter-riders. The training and fine tuning are completed using the IUPUI-CSRC-E-Scooter dataset which is available to the public as a benchmark dataset for e-scooter rider detection via http://escooterdataset.situated-intent.net. The details about the algorithm and the training and evaluation process can be found in [26].

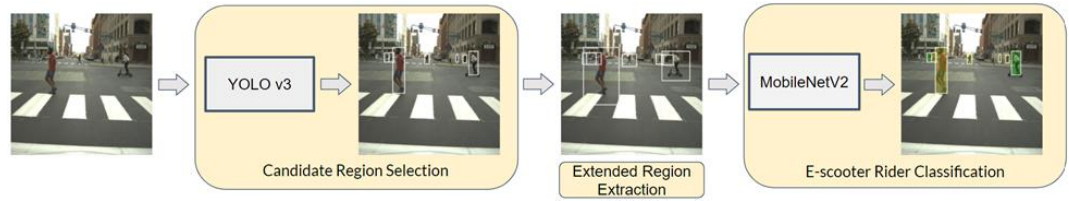

*Figure 4. E-scooter Detector.*

3. Local trajectory generation. Frame by frame, the relative positions of all the objects in the local coordinate system of the data collection platform are connected, filtered to remove outliers, interpolated to fill in missing data, and smoothened to reduce noise. Eventually, the local trajectories of these objects in the vehicle or e-scooter coordinate systems are generated.

4. The GPS-IMU fusion for ego-vehicle trajectory generation.

5. Global scene reconstruction. The global trajectories of the data collection platform and the local trajectories of all the surrounding objects relative to the data collection platform are fused. The reconstructed scene includes the global trajectories of the ego-car/ego-e-scooter and all the surrounding road users.

## DATA ANALYSIS

### Process
Cases with a time-to-collision (TTC) are defined as potential conflict cases. TTC indicates the potential crashes at a specific moment if both the ego vehicle and e-scooter maintain the current dynamic state (speed and heading direction). The complete data analysis process is explained in Figure 5.

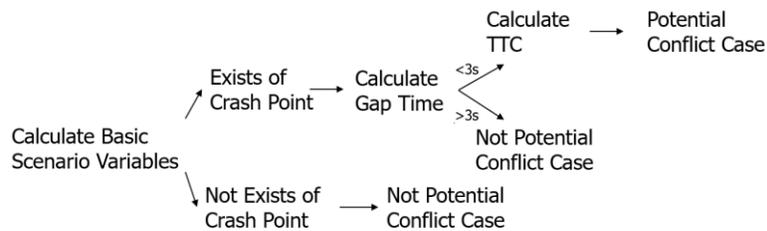

*Figure 5. Data Analysis Process.*

The minimum distance between the ego vehicle and e-scooter and their median speeds are first calculated. Based on the GPS coordinates of the current and previous points, the speed and heading angle are then calculated. Next, a crash point is determined in an intersection between the two trajectories. Finally, the cases with crash points are calculated for the gap time. A case with fewer than three (3) seconds of gap time is considered as a potential conflict case and used to compute the TTC. After the whole process, all cases are classified into non-potential and potential conflict cases. Moreover, from the vehicle-centric data the number of people wearing a backpack or helmet were also recorded.

### Scenario Variables Calculation
**Gap Time and Time-to-collision:** Given the E-scooter and vehicle's location, speed, and heading angle, they have potential conflicts if their coast trajectory intersect. If there is an intersection between the two coast trajectories, they must have a gap time, which is the time difference between their arrivals at the intersection. If the time difference between their arrival at the intersection is less than three seconds, it is defined as a possible crash. The definitions of gap time and time-to-collision are illustrated in Figure 6.

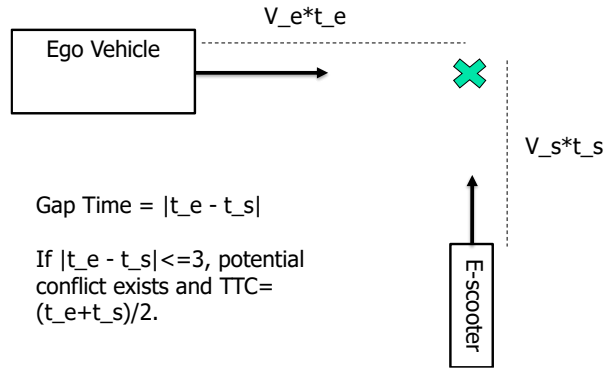

*Figure 6 Gap Time and Time-To-Collision Calculation.*

A threshold is set to filter out the gap time greater than 20 seconds. The main reason is that such a significant gap time indicates a low crash severity level but could contaminate the average gap time (by increasing it).

**Closest distance:** Since we have the synchronized GPS coordinates for both the e-scooter and vehicle from previous preprocessing, we calculate the distance difference between them at each timestamp. The distance difference is then sorted to find the closest difference.

**Median Speed:** The speed of either the vehicle or the e-scooter is calculated by dividing the distance difference between two consecutive GPS coordinates by the corresponding time difference. This gives the estimated speed for each timestamp. Finally, the median speed is calculated for the whole trajectory. The reason for using median speed over average speed is the ability to remove outliers or noises from compromising the calculation.

**RESULTS**

Scenario variables were calculated for all cases, potential conflicting cases, and cases in different countering geometries. A total of 182 e-scooter-vehicle encountering cases were analyzed from the e-scooter-centered data.

**Vehicle-Centered Data Analysis**
A total of 203 vehicle-e-scooter encounter cases were analyzed from the vehicle-centered data. Table 1 shows the scenario variables analyzed from the vehicle-centered data. Some key findings are as below:

1. The average shortest distances are around 15-16m and the minimum distance is around 3m

2. During encounters, the average vehicle speed is about 12mph, while that of an e-scooter is 8.5mph.

3. Some cases highlight the possibility of fatal crashes at high speeds, 30mph and 20mph, respectively for vehicles and e-scooters.

4. The average minimum TTC calculated for all the potential conflict cases is 3.77 seconds, with the minimum TTC of about 1 second. Some of these cases are associated with small proximity and higher risks.

5. The minimum gap time for all cases has an average value of 3.09 seconds. The two entities keep about 10-15m apart when there is a need to cross each other's paths.

*Table 1*
*Scenario Variables for all Vehicle-Centered Cases.*

| Variables | Average | Minimum | Maximum |
| --- | --- | --- | --- |
| Minimum distance (meters) | 15.29 | 3.14 | 61.82 |
| Ego-vehicle Median Speed (m/s) | 5. 3 | 0 | 14.61 |
| Ego-vehicle Median Speed (mph) | 11.86 | 0 | 32.68 |
| E-scooter Median Speed (m/s) | 3.79 | 0 | 8.99 |
| E-scooter Median Speed (mph) | 8.48 | 0 | 20.11 |
| Minimum TTC for Potential Conflict Cases (mTTC ) (seconds) | 3.77 | 0.96 | 13.85 |
| Minimum Gap Time (Seconds) | 3.09 | 0.03 | 16.91 |

**Analysis of Potential-Conflict Cases**: Among all the 203 encountering cases analyzed from the vehicle-centered data, 26.1% are potential conflict cases (Figure 7). Figure 8 shows the distribution of the minimum TTC for these cases. We have 10 cases in the range of 0-2 seconds, 28 cases in the range of 2-4 seconds, and the rest is more than 4 seconds.

Figure 9 shows the distribution of these potential conflict cases with the following defined risk levels:

- High risk: mTTC < 1 seconds
- Medium risk: mTTC < 2.5 seconds
- Low risk: mTTC >= 2.5 seconds

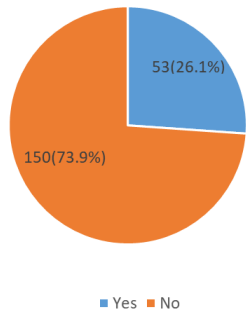

*Figure 7 Distributions of Potential Conflict and Non-Potential Conflict cases*

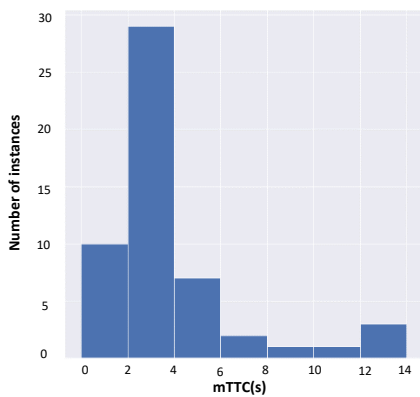

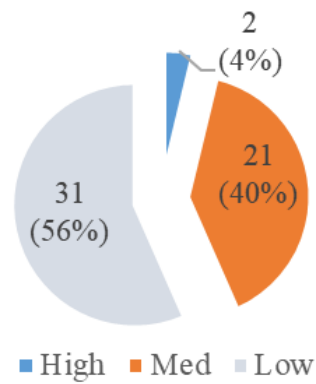

*Figure 8 Distribution of mTTC for Potential Conflict Cases*

*Figure 9 Distribution of Potential Conflict Cases with Different Risk Levels*

**Comparison Between Baseline and Potential-Conflict Cases:** Table 2 compares potential conflict cases and baseline (non-potential conflict) cases in the key scenario variables.

*Table 2*
*Comparison of Scenario Variables for Potential and Baseline Cases*

| Scenario Variables | Potential Conflict Cases | Baseline Cases |
|---|---|---|
| Average Minimum Distance | 16.22 m | 14.94 m |
| Average Ego-Vehicle Median Speed | 18.25 mph | 9.51 mph |
| Average E-scooter Median Speed | 8.43 mph | 8.61 mph |
| Average Minimum Gap Time | 0.77 s | 5.54 s |

The results show that the average minimum distance and the average e-scooter median speed among these two types of cases are similar. The average vehicle speed is much higher in the potential conflict cases at 18.25 mph, compared to 9.51 mph in the baseline cases. Not surprisingly, the average minimum gap time in the potential conflict cases is much shorter than the baseline cases. The potential conflict cases have a higher average minimum distance than the baseline cases. This is because, in baseline cases, the e-scooters riding on sidewalks tend to be near the driving lane and the coast trajectories do not intersect. In potential conflict cases, with some exception, the vehicle driver or the e-scooter rider tend to take necessary actions before their anticipated point of intersection to avoid a crash. The average median speed of the ego-vehicle is also much higher.

**Cases in Different Scenarios:** Table 3 shows the four encountering geometries analyzed in this project. All the cases are classified into these four geometries. If one case may be classified into multiple geometries, only one geometry is selected based on the main interaction phase. The results are shown in Figure 10. Most cases are parallel cases (about 73%), and the rest 27% are crossing cases, out of which more than half of the encounters (51%) are parallel cases with the same direction, and about 22% are parallel with opposite directions. E-scooter crossing from left cases stands for about 17% of all the cases, and the rest 9% are e-scooter crossing from right cases. Another important observation is that only 2-3% of e-scooter riders wear helmets when riding the e-scooters in the street.

*Table 3*
*Four Geometries for Car and E-scooter Encounters*

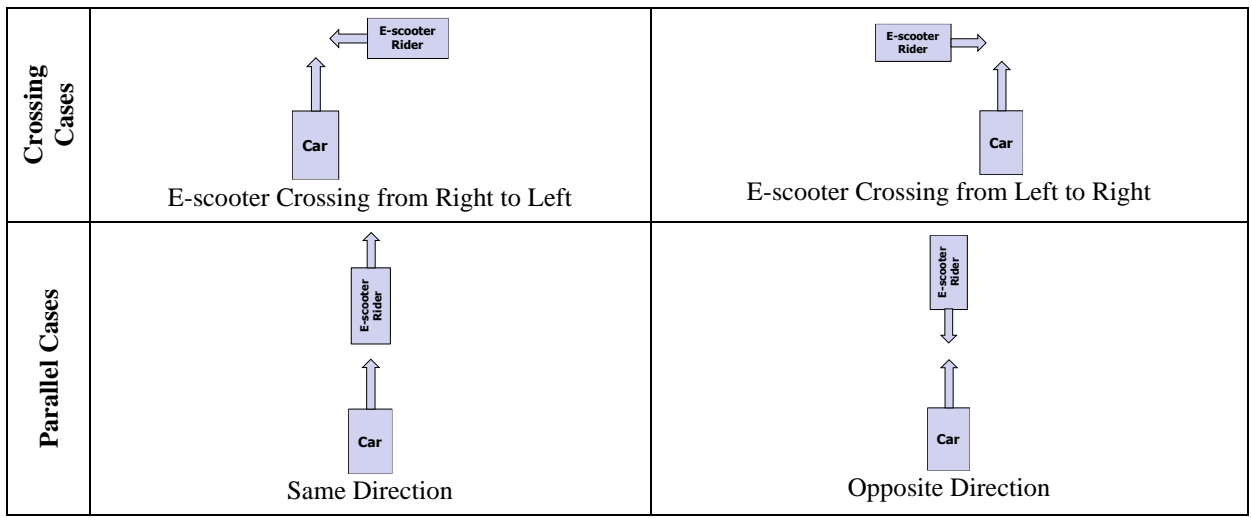

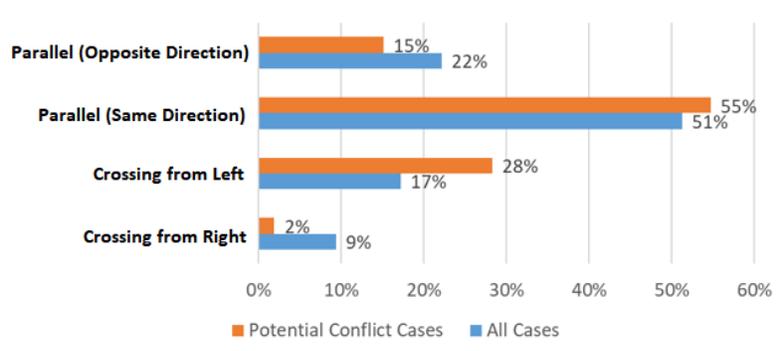

*Figure 10. Distribution of Geometry Types for All Cases and Potential Conflict Cases*

**E-Scooter-Centered Data Analysis**

A total of 182 e-scooter-vehicle encountering cases were analyzed from the e-scooter-centered data.

Table 4 shows the scenario variables for the e-scooter-centered data. Some of the key findings are:

1. Compared with the vehicle-centered data, the e-scooter-centered data can capture more cases with closer contact between cars and e-scooter riders.

2. Compared to the vehicle-centered data, the e-scooter-centered cases have higher speeds for both e-scooters and cars. This may be because most e-scooter-centered cases are collected at mid-block road locations, while many vehicle-centered cases happen at intersections.

3. The average minimum TTC calculated for all the potential conflict cases is 3.99 seconds, with the minimum TTC of about 1 second. All these numbers are like the numbers in the vehicle-centered analysis.

4. The minimum gap time for all cases has an average value of 2.26 seconds. This can illustrate the average time differences among vehicles and e-scooter riders to pass the common zones from the e-scooter-centered data.

*Table 4*
*Scenario variables for all e-scooter-centered cases in comparison to vehicle-centered data.*

|  | E-scooter-centered data | | | Vehicle-centered data | | |
|---|---|---|---|---|---|---|
| Scenario Variables | Average | Minimum | Maximum | Average | Minimum | Maximum |
| Minimum Distance (meters) | 7.27 | 2.35 | 21.68 | 15.29 | 3.14 | 61.82 |
| Ego-vehicle Median Speed (m/s) | 6.1 | 0 | 18.4 | 5. 3 | 0 | 14.61 |
| Ego-vehicle Median Speed (mph) | 13.65 | 0 | 41.16 | 11.86 | 0 | 32.68 |
| E-scooter Median Speed (m/s) | 4.8 | 0 | 10.60 | 3.79 | 0 | 8.99 |
| E-scooter Median Speed (mph) | 10.74 | 0 | 23.71 | 8.48 | 0 | 20.11 |
| mTTC (seconds) | 3.99 | 1.05 | 12.51 | 3.77 | 0.96 | 13.85 |
| Minimum Gap Time (Seconds) | 2.26 | 0.01 | 12.76 | 3.09 | 0.03 | 16.91 |

**Analysis of Potential-Conflict Cases**: Among all the 182 encountering cases analyzed from the e-scooter-centered data, 45.1% of them are potential conflict cases (Figure 11). Figure 12 shows the distribution of the minimum TTC for these cases. The minimum distance is considerably different between the vehicle-centric and e-scooter-centric data, with the e-scooter-based distance being significantly less than the vehicle-based distance. During the e-scooter-centric data collection, the subjects were advised to follow e-scooter riding rules in Indianapolis, which is to ride on the street, alongside other motorized vehicles (avoiding sidewalks wherever possible). During the vehicle-centric data collection, it was noticed that majority of e-scooter were ridden on the sidewalks. This explains why the minimum distance between the two is noticeably different.

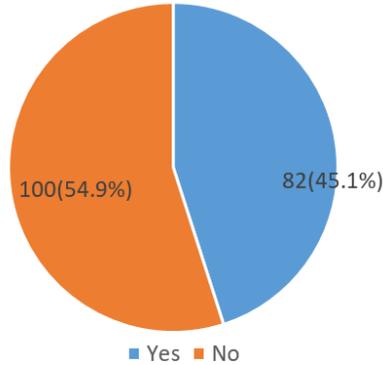

*Figure 11. Distributions of Potential Conflict and Non-Potential Conflict cases*

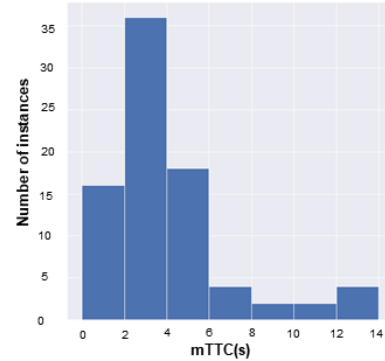

*Figure 12. Distribution of mTTC for Potential Conflict Cases in E-scooter-centered Data*

**Comparison Between Baseline and Potential-Conflict Cases:** Table 5 shows the comparison between potential conflict cases and baseline (non-potential conflict) cases e-scooter-centered data. One difference from the vehicle-centered data is that the average vehicle speed is comparable between potential-conflict and non-potential-conflict cases. The average e-scooter median speed is higher in potential conflict cases. This explains the difference in average minimum distance.

*Table 5*
*Comparison of scenario variables between potential-conflict and baseline cases in the e-scooter-centered data*

|  | **Potential Conflict Cases** | **Baseline Cases** |
|---|---|---|
| Average Minimum Distance | 7.67 m | 6.95 m |
| Average Ego-Vehicle Median Speed | 13.60 mph | 13.65 mph |
| Average E-scooter Median Speed | 11.63 mph | 10.07 mph |
| Average Minimum Gap Time | 0.84 s | 3.97 s |

**Cases in Different Scenarios:** The results for the different geometries are shown in Figure 13. For e-scooter-centered data, most of them are parallel cases (about 75%), and the rest of them (25%) are crossing cases. More than half of the encounters (52%) are parallel cases with the same direction, and about 23% are parallel with opposite directions. E-scooter crossing from left side stand for about 9% of all the cases, and the rest 16% are e-scooter crossing from right side. These findings are quite similar as the results from the vehicle-centered data analysis. The only difference is that there are more crossing-from-right cases than crossing-from-left cases in the e-scooter-centered dataset.

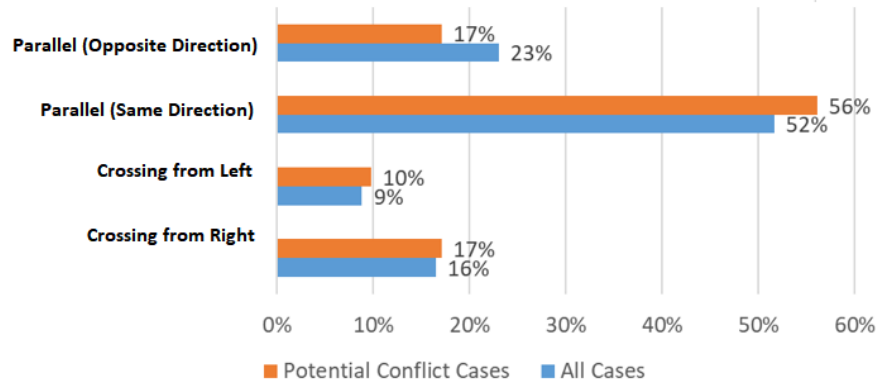

*Figure 13. Distribution of Geometry Types for All Cases and Potential Conflict Cases*

**CONCLUSIONS**

In this work, two data acquisition systems were developed to collect naturalistic data for vehicle-e-scooter-rider interactions. After carefully designing the routes and areas for vehicle-centered and e-scooter-centered data collection activities in the Indianapolis downtown area and IUPUI campus, 55 subjects with diverse demographic backgrounds were recruited to participate in the data collection experiments. 68 vehicle-centered data collection sessions and 14 e-scooter-centered data collection sessions were completed. To process the large-scale data collection, an efficient data preprocessing pipeline was developed. A state-of-the-art algorithm was adopted to detect and track vehicles and humans from the collected video data. Furthermore, a computer-vision approach was developed with a new benchmark dataset to detect e-scooter riders efficiently and automatically.

The study has shown the behavior metrics of e-scooter riders in different vehicle-e-scooter interactions. E-scooter rider detection benchmark dataset and classifier are also released publicly. The results of this study will help to fill the gap in quantitative empirical data about e-scooter rider behaviors, which can potentially support the development and testing of vehicle-e-scooter collision avoidance mitigation systems. In the follow up effort, the results can be further integrated with crash data to better model e-scooter crash risks and guide the training of e-scooter behavior prediction algorithms with more labels to the dataset.

**REFERENCES**

[1]  Howe, E. and Bock, B., 2018. Global Scooter Sharing Market Report.
[2]  National Association of City Transportation Officials (NACTO), 2020, Shared Micromobility in the U.S.: 2019. Available online: https://nacto.org/shared-micromobility-2019/ (Accessed February 2021).
[3]  Almannaa, M.H., Alsahhaf, F.A., Ashqar, H.I., Elhenawy, M., Masoud, M. and Rakotonirainy, A., 2021. Perception Analysis of E-Scooter Riders and Non-Riders in Riyadh, Saudi Arabia: Survey Outputs. Sustainability, 13(2), p.863.
[4]  Ching-Fu, C., Cheng-Chien, K. and Yi-Ju, L., 2017, January. Investigating barriers and facilitators of attitude and intention to use e-scooter sharing system. In 22nd International Conference of Hong Kong Society for Transportation Studies: Transport and Society, HKSTS 2017 (pp. 399-406). Hong Kong Society for Transportation Studies Limited.
[5]  Aguilera-García, Á., Gomez, J. and Sobrino, N., 2020. Exploring the adoption of moped scooter-sharing systems in Spanish urban areas. Cities, 96, p.102424.
[6]  Fitt, H. and Curl, A., 2019. Perceptions and experiences of Lime scooters: Summary survey results.
[7]  Huang, F.H. and Lin, S.R., 2018, August. A Survey of User Experience of Two-Wheeler Users in Long-Term Interactions. In Congress of the International Ergonomics Association (pp. 1465-1472). Springer, Cham.

[8] James, O., Swiderski, J.I., Hicks, J., Teoman, D. and Buehler, R., 2019. Pedestrians and e-scooters: An initial look at e-scooter parking and perceptions by riders and non-riders. Sustainability, 11(20), p.5591.

[9] Sanders, R.L., Branion-Calles, M. and Nelson, T.A., 2020. To scoot or not to scoot: Findings from a recent survey about the benefits and barriers of using E-scooters for riders and non-riders. Transportation Research Part A: Policy and Practice, 139, pp.217-227.

[10] Austin Public Health, 2018, Dockless Electric Scooter-Related Injuries Study. Report. Austin Public Health (APH), Available online via (accessed February 2021): https://www.austintexas.gov/sites/default/files/files/Health/Epidemiology/APH_Dockless_Electric_Scooter_Study_5-2-19.pdf

[11] Multnomah County Health Department, 2019. Scooter-related Injuries in Multnomah County July--November 2018. Online via (Accessed February 2021): https://www.portlandoregon.gov/transportation/article/709715.

[12] Alison Griswold, 2020, "At least 29 people have died in electric scooter crashes since 2018", Quartz report, accessed 02.2021. https://qz.com/1793164/at-least-29-people-have-died-in-electric-scooter-crashes/.

[13] Yang, H., Ma, Q., Wang, Z., Cai, Q., Xie, K. and Yang, D., 2020. Safety of micro-mobility: analysis of E-Scooter crashes by mining news reports. *Accident Analysis & Prevention*, *143*, p.105608.

[14] FHWA, 2017. National Household Travel Survey. Available online at: https://nhts.ornl.gov/vehicle-trips. Accessed (02.2021).

[15] Beck, S., Barker, L., Chan, A. and Stanbridge, S., 2020. Emergency department impact following the introduction of an electric scooter sharing service. *Emergency Medicine Australasia*, *32*(3), pp.409-415.

[16] Badeau, A., Carman, C., Newman, M., Steenblik, J., Carlson, M. and Madsen, T., 2019. Emergency department visits for electric scooter-related injuries after introduction of an urban rental program. *The American journal of emergency medicine*, *37*(8), pp.1531-1533.

[17] Ma, Q., Yang, H., Mayhue, A., Sun, Y., Huang, Z. and Ma, Y., 2021. E-scooter safety: the riding risk analysis based on mobile sensing data. *Accident Analysis & Prevention*, *151*, p.105954.

[18] Dozza, M., Piccinini, G.F.B. and Werneke, J., 2016. Using naturalistic data to assess e-cyclist behavior. *Transportation research part F: traffic psychology and behaviour*, *41*, pp.217-226.

[19] Prabu, A., Ranjan, N., Li, L., Tian, R., Chien, S., Chen, Y., and Sherony, R., 2022, SceNDD: A Scenario-based Naturalistic Driving Dataset, *IEEE 25th International Conference on Intelligent Transportation Systems (ITSC)*, pp. 4363-4368, doi: 10.1109/ITSC55140.2022.9921953.

[20] Prabu, A., Shen, Dan., Tian, R., Chien, S., Li, Lingxi., Chen, Y., and Sherony, R., 2021. A wearable data collection system for studying micro-level e-scooter behavior in naturalistic road environment. *Proc. Fast-zero'21,* Kanazawa, Japan.

[21] De Alvis, C., Shan, M., Worrall, S., and Nebot, E., 2019, May. Uncertainty Estimation for Projecting Lidar Points onto Camera Images for Moving Platforms. In 2019 International Conference on Robotics and Automation (ICRA) (pp. 6637-6643). IEEE.

[22] Zhao, W., Chellappa, R., Phillips, PJ., and Rosenfeld, A., 2003. Face recognition: A literature survey. Acm Computing Surveys (CSUR), 35(4):399–458.

[23] Zhou, X., Koltun, V. and Krähenbühl, P., 2020, August. Tracking objects as points. In *European Conference on Computer Vision* (pp. 474-490). Springer, Cham.

[24] Redmon, J. and Farhadi, A., 2018. Yolov3: An incremental improvement. arXiv preprint arXiv:1804.02767.

[25] Sandler, M., Howard, A., Zhu, M., Zhmoginov, A. and Chen, L.C., 2018. Mobilenetv2: Inverted residuals and linear bottlenecks. In Proceedings of the IEEE conference on computer vision and pattern recognition (pp. 4510-4520).

[26] Apurv, K., Tian, R. and Sherony, R., 2021. Detection of E-scooter Riders in Naturalistic Scenes. arXiv preprint arXiv:2111.14060.